\documentclass{llncs}

\usepackage{graphicx} 
\usepackage{mathtools} 
\usepackage{caption} 
\usepackage{subcaption} 
\usepackage{multirow} 
\usepackage{float} 
\usepackage{amsmath} 
\usepackage{listings} 
\usepackage{xcolor} 

\newcommand{\cmnt}[1]{\ignorespaces}

\begin{document}

\title{Appliance of network theory in economic geography}
\titlerunning{Appliance of network theory in economic geography}  
%
\author{Alexandra Barina\inst{1,3}, Gabriel Barina\inst{1} and Mihai Udrescu\inst{1,2}}
%
%
%
\institute{Department of Computer and Information Technology, Politehnica University of Timi\c{s}oara, Timi\c{s}oara, Romania\\
\and
Timi\c{s}oara Institute of Complex Systems, Timi\c{s}oara, Romania
\and
Corresponding author at Faculty of Automation and Computers, University Politehnica of Timisoara, Bd. Vasile Parvan 2, 300223, Timisoara, Romania, email: vadasanmaria@gmail.com
}

\maketitle              

\begin{abstract}
A continuously evolving geography requires a good understanding in networks. As such, this paper accounts for theories and applications of complex networks and their role both in geography in general, as well as in determining various geographical network trajectories. It assesses how links between agents lead to an evolutionary process of network retention, as well as network variation, and how geography influences these mechanisms.

\end{abstract}

\section{Introduction}

The limited insight offered by the neoclassical growth theory \cite{solow1999neoclassical} pertaining both regional growth \cite{salaxx1996regcohes,gallofing2019reggrowth} and geographical agglomeration \cite{fuji2003doesgeoaglo,bertdecrop2005geoaglo,dup2007dogeoaglo,stel2005wherecityform} has brought forth completely new approaches to economic geography \cite{boschmalamby1999evolecgeo,boschmafrenk2017whyecgeonot}. Integrating both growth theories and endogenous explanations pertaining to regional economy development \cite{frenkbosch2007frmwrkurb,martsun2006pathdepend,kog2015theoretprogrss}, valuable insight is thus gained by economic geographers. Shifting their attention towards the understanding of geographical proximity of both technological and economical innovations \cite{weiluz2016endotecchg}, geographers are interested in understanding how innovation is made and why innovative practice often concentrates in certain geographical proximities \cite{stel2005wherecityform}.

As a result, one way of analyzing certain regional economical developments \cite{storp1997regionalworld} is to apply network science and identify certain interactions in generated economic network. As such, geographers -- especially the ones from the economic department -- have become gradually receptive to network science, due to its various applications in analyzing economic outcomes \cite{gvet2005socstruc,peliz2018socmobility}. Just like evolutionary economics, network theory uses regional clusters to study various conditions and outcomes in a dynamic fashion \cite{diez2018locknow}. In the dynamic field of geography, networks have been applied with success over the course of many years already and have coined specific terminologies pertaining geographical clusters and even globalization.

Following a critical review published by economic geographer Gernot Grabher \cite{grabher2006trading}, it has come to the attention however, that the application of network science in the field of geography has been done either superficially or selectively. As such, in this paper we address the development of network science and its application in geography. Using emerging research on network evolution \cite{ferg2017hypconn,knud2004selectheor}, we aim at integrating concepts into the field of economic geography. As a result, we propose the following objectives: 
\begin{itemize}
  \item Define a geographical network trajectory \cite{bofrenk2018evolecgeo} which incorporates the application of network science in order to study the evolution of networks across regions.
  \item Analyze retention-, and variation-mechanisms in network structures which are endogenous to network evolution.
  \item  Determine models of variation in network trajectories by taking into consideration various regional innovations.
\end{itemize}

The purpose of this paper is to discuss the contribution of network theory to an evolutionary economic geography, as well as to prove that innovation and regional growth results from the bridging of several (unconnected) networks. As such, after addressing basic concepts in section \ref{section2}, section \ref{section3} presents the existing mechanisms responsible for selection-, retention-, and variation-mechanisms. Conclusively, section \ref{section_conclusion} ends this paper with a general discussion pertaining sources of innovation for regional growth.

\section{Theoretical Background}
\label{section2}

In this section we offer theoretical background pertaining social network analysis and its application in geography, with particular focus on the economic side of it.

\subsection{Social Network Analysis}

Social network analysis is the process of investigating various social structures  by using networks and graph theory \cite{kimhas2018socnetwanalys}. As such, a social network is a construction formed by individuals (actors, agents) with bidirectional connections (relationships, friendships) between them, effectively resembling a real-world structure of our own society. Social networks derive from complex networks, together with which they form part of the nascent field of network science \cite{bormeh2009netwanalsocscienc,easklein2010netwcrowds}, ever since their first introduction to literature in the 1970's \cite{kenperc2015netofnetw}. Based on graph theory and network theory, empirical observations of real-world networks and sociology, their main purpose is to model the various relationships from our own society.

Underpinned by empirical studies of real-world systems, complex networks have gained increased interest due to their applicability in various social \cite{topirceanu2016tolerance,duma2014motifs,top2014geneticalfbk} and scientific fields, such as biology, economy \cite{gab2014musenet} and geography \cite{ballrig2017geocomplexknowge}. For instance, in the economic area it is highly important to understand the mechanism in which certain economic agents handle better than others \cite{jack2008average}.

\subsection{Geography}

The field of geography can be linked to complex networks by factoring in two main properties, namely proximity and place, both very important views of geography when taking into consideration the concept of geographical network trajectory.

Proximity, a latent effect of various economic processes, influences network formation in a direct manner; as such, the most widely used approach in economic geography aims at addressing this issue. It is directly influenced by two underlying social technologies present throughout all existing relations of economic geography, namely communication technology \cite{storp2004buzz} and transport technology \cite{marquis2003pressurepast}. As such, only by analyzing communication preferences of social actors and mobility opportunities can we also determine the existing relation between physical space and economic interaction; in other words, constraints of proximity only apply if face-to-face is the only means of communication while travel is denied.

By using the notion of resource bundle from the theory of the growth \cite{penrose1959growthfirm}, a place -- the second property by which we can establish a link between networks and geography -- can be considered both a resource and opportunity. As such, a place-specific resource offers a source of contextuality, difference and contingency for economic development \cite{say1991behind,batgluck2005resources}. Conclusively, by taking into consideration structural (\textit{e.g.,} social capital, structural holes, etc.), material, social and economical resources, places do not coerce the formation of a given network, but various social interactions do influence the geography (of the network) \cite{frenkbosch2007frmwrkurb}.

\subsection{Geographical network trajectory}

The main focus of any theory pertaining network evolution is how structural layout of a given network at time 1 affects the interactions among agents (nodes) and/or the formation of new relationships (ties) at time 2 \cite{kenkno2002interorganizat}. As such, the concept that combines network, geography and evolution in order to analyze the above scenario is the geographical network trajectory \cite{kiltsai2003socntworganiz}. It describes from a geographic point of view the formation and dissolution of relationships (ties) within a given network. This perspective effectively shifts the analysis of single relations to the analysis of the overall relations from within a network. As such, a theory behind network evolution takes into consideration not only the appearance of a new tie, but also the impact that the system imposes on the creation of additional relationships.

\section{The evolutionary process in geographical network trajectory}
\label{section3}

A given evolutionary system can be defined by the following principles: selection, retention as well as variation \cite{nelswint2002evoltheorizing,hodgslamb2018pastfuture}. As such, in the following sub-sections we present these principles in the context of network theory.

\subsection{Selection}

The selection mechanism is based on environment. In the field of economics, for instance, it is the market competition that dictates the existence and layout of various producers \cite{knud2002selectheory}. In field of social network analysis however,  selection refers to the presence of relationships (ties) between agents (nodes) of the same network \cite{gulati1995famtrust,stu1998netwpositions,gulgarg1999where,ahuja2000collab,venklee2004preflinkage,sole2002selection}. This selection is the result not only of an external environment, but also of the various decisions made by the mutual agents linked together. This implies that the selection mechanism is based on both the requirement of constant adaptation, as well on a strategy focused on improving the relations between both agents, to the point of mutual benefit.

In all economical networks new relationships occur either between two producers who already have a shared history, or new producers, without any previous relationships. As such, a theory of network evolution must take into consideration both the emergence of dynamic links, as well as possible disappearance of these; both cases imply the presence of another factor, namely cost. 

Creating and/or maintaining relations is a costly resource, and is greatly dependent on each producer as a limiting factor. As such, producers analyze the presence of relations mainly from a utility perspective, and how they would benefit from it. One such benefit would be the possibility of accessing external resources \cite{pfefsala1978externcontrol}; this, however, would also automatically decrease the producers' attractiveness for new relations. In this case, tie selection can be defined as the competitive allocation of limited relationships where the presence of such a relationship automatically implies possible lost opportunities. This suggests that tie selection is a competitive process which hinges on both exogenous changes (\textit{e.g.,} regulatory markets) \cite{derrkec2013realeffects} as well as endogenous dynamics (\textit{e.g.,} one producer has become more attractive
because of its alliance with a third party) \cite{hallghil2008natdisaster,galcut2007neighbour}. In what follows, we further address these two mechanisms, with focus on endogenous mechanisms of network evolution like retention and variation of network structures.

\subsection{Retention}

The retention mechanism of a network refers to the inclination towards a given (future) tie selection based on past decisions and their outcome. This mechanism stems from either persistent (\textit{i.e.,} slow-decaying) or new, path-dependent ties. According to sociologist Ronald S. Burt, the decay is a power function of time, in which decay probability decreases the older a tie and/or node is \cite{burt2000decayfunc}. However, as mentioned before, retention also comes from the appearance of path-dependent ties. As such, while studies on tie-decays analyze the life-span of existing relationships, another approach is to analyze where the next tie is most likely to form, even without (significant) change in layout, centrality or density distributions or even fragmentation \cite{vanamani2014pathdepend}. As such, the existing literature on the retention mechanism describes three main theories, described in what follows.

\subsubsection{Preferential attachment}

This theory states that when creating new ties in a scale-free network topology \cite{barab2002evolsocnetw}, agents with a high degree are capable of increasing their degree even more and at a much faster pace, while agents with a small degree will inevitably stagnate in the process \cite{barabrek1999emergence,wangchen2003complexnetw}. From this, we can deduce that well-positioned agents have a clear, accumulative advantage (over time) compared to the un-connected and peripheral agents \cite{gulgarg1999where,powkop1996interorg}. However, in a real-life scenario, this does have a limitation, in that agents can only maintain a limited number of relationships, and as such, the process of centralization is, in fact, empirically finite \cite{holmeedling2004structime}.

\subsubsection{Embedding}

This theory suggests that new relationships are formed based on the trust of the old ones. The increased interconnection to which this leads is the process of social embedding \cite{gulgarg1999where}. As a result, this leads to persistent network structures, as well as the formation of new mentalities within groups of interconnected agents \cite{baum2003wheresmworlds}. 

\subsubsection{Multi-connectivity}

This theory pertains to the fact that networks expend by means of a process meant to induce diversity. Creating multiple independent ties between them, agents gain a cumulative advantage of multi-connectivity, and become more attractive for the formation of new relationships \cite{powwhite2005netwdynam}. As these new links form, they reinforce clusters of agents, who become more cohesive over time. This led Walker \cite{walkkog1997soccapital} et. al. to argue in favor of path-dependence in network growth \cite{hite2001evolfirm,cookbosch2011handbook}.

As a whole, the previously mentioned theories (\textit{i.e.,} preferential attachment, embedding and multi-connectivity) represent a cumulative path-dependent retention mechanisms found in networks of various types. Additionally, geographical location also affects the evolution of the network trajectory \cite{stu1998netwpositions,gluck2007econgeogr,sorestu2001syndicat,powkop2002spatclustering}; however, the impact is only limited to a certain regional level, where face-to-face communication is allowed \cite{powwhite2005netwdynam}. New relationships are more likely to form when two agents are located in the same region (\textit{i.e.,} co-located), even if the involved agents are not central within a given network \cite{owpow2004knownetw,schutstam2003evolnatyoungfirm}. As a result, there is numerous evidence for retention mechanisms in networks from a geographical point of view. 

Complex networks successfully convey important aspects of cumulative, path-dependent evolution over time between individual agents as well as between groups of agents (clusters). However, these retention mechanisms may also lead to a situation known as local lock-in \cite{martsun2006pathdepend,has2005unlockregecon}, in which certain already-existing patterns prohibit the appearance of new ones. Lock-ins, however, can be overcome by the emergence of new variation.

\subsection{Variation}

As we can see, variation is the result of endogenous network mechanisms for both tie formation and dissolution. In a network, variation opposes an existing network trajectory in favor of selection of new ties. As such, even though variation refers to changes in network structure, it is defined at the level of tie selection, which in turn, affects the re-organization of the network \cite{burt2004structholes}. Empirical data has shown that creating and maintaining an active relationship with an agent outside of ones own network would shorten the given agents membership duration \cite{mcphers1992socnetworks}; it is due to this phenomenon that variations happen in social and economic networks, which countervail the previously-mentioned retention mechanisms. Then again, once such a link is established, the cumulative retention mechanisms (\textit{i.e.,} preferential attachment, embedding and multi-connectivity) create new opportunities for new links, thus effectively countering existing patterns of path-dependence \cite{baum2003wheresmworlds,row2005timebreakup}.


\section{Conclusions}
\label{section_conclusion}

Evolutionary economic networks are greatly influenced by retention-, and variation-mechanisms which, in turn, create path-dependent network trajectories \cite{martsun2006pathdepend} by implicating certain path-dependence attributes (\textit{e.g.,} preferential attachment, embeddedness, multi-connectivity, \textit{etc.}) \cite{gluck2007econgeogr}; moreover, evolutionary theory of economic growth must also be able to explain innovation in regional economic development. However, since evolutionary theory in fields like economy \cite{boschmalamby1999evolecgeo}, sociology \cite{dietz1990evolutionary}, geography \cite{boschmalamby1999evolecgeo,boschmart2007constructing} is still in its infancy, the observations drawn in this paper are exploratory only. 

As a result of limited application of evolutionary theory, geographers have turned to network theory from the new domain of social network analysis. As such, rather than offering a coherent theory pertaining regional evolution, network theory offers a new, evolutionary perspective to economic geography. Offering a way of promoting economic growth and innovation, it also represents a means to analyze the social values and economic interests between regional structures and individuals.

The analysis of network evolution, including from an economic point of view, is important due to the emergence or disappearance of both agents (nodes) and relationships (ties) simultaneously. As such, when new relationships form, the causes and consequences of network growth must also be addressed: what is the relationship between a path-dependent network trajectory and the growth rate of a network? Apart from this simple question, there are many other questions (economic) geographers need answers to, however an evolutionary analysis of changes within a network requires not only relational-, but also longitudinal data spanning over extensive time-periods \cite{baum2003wheresmworlds}. Seeing that this kind of data can only be analyzed using network theory, it offers a perfect incentive for geographers to apply methodologies offered only by social network analysis.



\begin{thebibliography}{99}

\bibliography{biblio}

\bibitem{solow1999neoclassical}Solow, R.M., 1999. Neoclassical growth theory. Handbook of macroeconomics, 1, pp.637-667.

\bibitem{salaxx1996regcohes}Sala-i-Martin, X.X., 1996. Regional cohesion: evidence and theories of regional growth and convergence. European Economic Review, 40(6), pp.1325-1352.

\bibitem{gallofing2019reggrowth}Le Gallo, J. and Fingleton, B., 2019. Regional growth and convergence empirics. Handbook of regional science, pp.1-28.

\bibitem{fuji2003doesgeoaglo}Fujita, M. and Thisse, J.F., 2003. Does geographical agglomeration foster economic growth? And who gains and loses from it?. The Japanese Economic Review, 54(2), pp.121-145.

\bibitem{bertdecrop2005geoaglo}Bertinelli, L. and Decrop, J., 2005. Geographical agglomeration: Ellison and Glaeser's index applied to the case of Belgian manufacturing industry. Regional Studies, 39(5), pp.567-583.

\bibitem{dup2007dogeoaglo}Dupont, V., 2007. Do geographical agglomeration, growth and equity conflict?. Papers in Regional Science, 86(2), pp.193-213.

\bibitem{stel2005wherecityform}Stelder, D., 2005. Where do cities form? A geographical agglomeration model for Europe. Journal of Regional Science, 45(4), pp.657-679.

\bibitem{boschmalamby1999evolecgeo}Boschma, R.A. and Lambooy, J.G., 1999. Evolutionary economics and economic geography. Journal of evolutionary economics, 9(4), pp.411-429.

\bibitem{boschmafrenk2017whyecgeonot}Boschma, R.A. and Frenken, K., 2017. Why is economic geography not an evolutionary science? Towards an evolutionary economic geography. In Economy (pp. 127-156). Routledge.

\bibitem{frenkbosch2007frmwrkurb}Frenken, K. and Boschma, R.A., 2007. A theoretical framework for evolutionary economic geography: industrial dynamics and urban growth as a branching process. Journal of economic geography, 7(5), pp.635-649.

\bibitem{martsun2006pathdepend}Martin, R. and Sunley, P., 2006. Path dependence and regional economic evolution. Journal of economic geography, 6(4), pp.395-437.

\bibitem{kog2015theoretprogrss}Kogler, D.F., 2015. Evolutionary economic geography–Theoretical and empirical progress.

\bibitem{weiluz2016endotecchg}Wiebe, K.S. and Lutz, C., 2016. Endogenous technological change and the policy mix in renewable power generation. Renewable and Sustainable Energy Reviews, 60, pp.739-751.

\bibitem{storp1997regionalworld}Storper, M., 1997. The regional world: territorial development in a global economy. Guilford press.

\bibitem{gvet2005socstruc}Granovetter, M., 2005. The impact of social structure on economic outcomes. Journal of economic perspectives, 19(1), pp.33-50.

\bibitem{peliz2018socmobility}Güell, M., Pellizzari, M., Pica, G. and Rodríguez Mora, J.V., 2018. Correlating social mobility and economic outcomes. The Economic Journal, 128(612), pp.F353-F403.

\bibitem{diez2018locknow}Díez-Vial, I. and Montoro-Sánchez, Á., 2018. How Local Knowledge Networks and Firm Internal Characteristics Evolve Across Time Inside Science Parks. In Agglomeration and Firm Performance (pp. 139-153). Springer, Cham.

\bibitem{grabher2006trading}Grabher, G., 2006. Trading routes, bypasses, and risky intersections: mapping the travels ofnetworks' between economic sociology and economic geography. Progress in human geography, 30(2), pp.163-189.

\bibitem{ferg2017hypconn}Ferguson, N., 2017. The False Prophecy of Hyperconnection: How to Survive the Networked Age. Foreign Aff., 96, p.68.
\bibitem{knud2004selectheor}Knudsen, T.R., 2004. General selection theory and economic evolution: The price equation and the replicator/interactor distinction. Journal of Economic Methodology, 11(2), pp.147-173.

\bibitem{bofrenk2018evolecgeo}Boschma, R. and Frenken, K., 2018. Evolutionary economic geography (pp. 213-229). Oxford: Oxford University Press.

\bibitem{kimhas2018socnetwanalys}Kim, J. and Hastak, M., 2018. Social network analysis: Characteristics of online social networks after a disaster. International Journal of Information Management, 38(1), pp.86-96.

\bibitem{bormeh2009netwanalsocscienc}Borgatti, S.P., Mehra, A., Brass, D.J. and Labianca, G., 2009. Network analysis in the social sciences. science, 323(5916), pp.892-895.

\bibitem{easklein2010netwcrowds}Easley, D. and Kleinberg, J., 2010. Networks, crowds, and markets (Vol. 8). Cambridge: Cambridge university press.

\bibitem{kenperc2015netofnetw}Kenett, D.Y., Perc, M. and Boccaletti, S., 2015. Networks of networks–An introduction. Chaos, Solitons \& Fractals, 80, pp.1-6.

\bibitem{topirceanu2016tolerance} Topirceanu, A., et al. "Tolerance-based interaction: A new model targeting opinion formation and diffusion in social networks." PeerJ Computer Science 2, 2016

\bibitem{duma2014motifs} Duma, A., and Topirceanu, A. "A network motif based approach for classifying online social networks." Applied computational intelligence and informatics (SACI), 2014 IEEE 9th international symposium on, pp. 311-315. IEEE, 2014

\bibitem{top2014geneticalfbk}Topirceanu, A., Udrescu, M. and Vladutiu, M., 2014. Genetically optimized realistic social network topology inspired by facebook. In Online Social Media Analysis and Visualization (pp. 163-179). Springer, Cham.

\bibitem{gab2014musenet}Topirceanu, A., Barina, G. and Udrescu, M., 2014, September. Musenet: Collaboration in the music artists industry. In 2014 European Network Intelligence Conference (pp. 89-94). IEEE.
\bibitem{ballrig2017geocomplexknowge}Balland, P.A. and Rigby, D., 2017. The geography of complex knowledge. Economic Geography, 93(1), pp.1-23.

\bibitem{jack2008average}Jackson, M.O., 2008, December. Average distance, diameter, and clustering in social networks with homophily. In International Workshop on Internet and Network Economics (pp. 4-11). Springer, Berlin, Heidelberg.

\bibitem{storp2004buzz}Storper, M. and Venables, A.J., 2004. Buzz: face-to-face contact and the urban economy. Journal of economic geography, 4(4), pp.351-370.
\bibitem{marquis2003pressurepast}Marquis, C. (2003) The pressure of the past: Network imprinting in intercorporate communities. Administrative Science Quarterly, 48: 655-89. Marsden, P. V. (1990) Network data and measurement. Annual Review of Sociology, 16: 435-63.

\bibitem{penrose1959growthfirm}Penrose, E., 1959. The theory of the growth of the firm. JohnW iley \& Sons, New York.

\bibitem{say1991behind}Sayer, A. (1991) Behind the locality debate: Deconstructing geography's dualisms. Environment and Planning A, 23: 283-308.

\bibitem{batgluck2005resources}Bathelt, H. and Glückler, J., 2005. Resources in economic geography: from substantive concepts towards a relational perspective. Environment and Planning A, 37(9), pp.1545-1563.
\bibitem{kenkno2002interorganizat}Kenis, P. and Knoke, D., 2002. How organizational field networks shape interorganizational tie-formation rates. Academy of Management Review, 27(2), pp.275-293.
\bibitem{kiltsai2003socntworganiz}Kilduff, M., Tsai, W. (2003) Social Networks and Organizations. London: Sage. 

\bibitem{nelswint2002evoltheorizing}Nelson, R. R., Winter, S. G. (2002) Evolutionary theorizing in economics. Journal of Economic Perspectives, 16: 23-46.

\bibitem{hodgslamb2018pastfuture}Hodgson, G.M. and Lamberg, J.A., 2018. The past and future of evolutionary economics: some reflections based on new bibliometric evidence. Evolutionary and Institutional Economics Review, 15(1), pp.167-187.

\bibitem{knud2002selectheory}Knudsen, T. (2002) Economic selection theory. Journal of Evolutionary Economics, 12: 443-70.

\bibitem{gulati1995famtrust}Gulati, R. (1995) Does familiarity breed trust? The implications of repeated ties for contractual choice in alliances. Academy of Management Journal, 38: 85-112.
\bibitem{stu1998netwpositions}Stuart, T. E. (1998) Network positions and propensities to collaborate: an investigation of strategic alliance formation in a high-technology industry. Administrative Science Quarterly, 43: 668-98.

\bibitem{gulgarg1999where}Gulati, R., Gargiulo, M. (1999) Where do interorganizational networks come from? American Journal of Sociology, 104: 1439-93.

\bibitem{ahuja2000collab}Ahuja, G. (2000) Collaboration networks, structural holes, and innovation: A
longitudinal study. Administrative Science Quarterly, 45: 425-55.

\bibitem{venklee2004preflinkage}Venkatraman, N., Lee, C.-H. (2004) Preferential linkage and network evolution: a conceptual model and empirical test in the US video game sector. Academy of Management Journal, 47: 876-92.
\bibitem{sole2002selection}Solé, R.V., Ferrer‐Cancho, R., Montoya, J.M. and Valverde, S., 2002. Selection, tinkering, and emergence in complex networks. Complexity, 8(1), pp.20-33.

\bibitem{pfefsala1978externcontrol}Pfeffer, J., Salancik, G. R. (1978) The External Control of Organizations. New York: Harper and Row.

\bibitem{derrkec2013realeffects}Derrien, F. and Kecskés, A., 2013. The real effects of financial shocks: Evidence from exogenous changes in analyst coverage. The Journal of Finance, 68(4), pp.1407-1440

\bibitem{hallghil2008natdisaster}Hallegatte, S. and Ghil, M., 2008. Natural disasters impacting a macroeconomic model with endogenous dynamics. Ecological Economics, 68(1-2), pp.582-592.

\bibitem{galcut2007neighbour}Galster, G., Cutsinger, J. and Lim, U., 2007. Are neighbourhoods self-stabilising? Exploring endogenous dynamics. Urban Studies, 44(1), pp.167-185.

\bibitem{burt2000decayfunc}Burt, R. (2000) Decay functions. Social Networks, 22: 1-28.

\bibitem{vanamani2014pathdepend}Vanacker, T., Manigart, S. and Meuleman, M., 2014. Path‐dependent evolution versus intentional management of investment ties in science‐based entrepreneurial firms. Entrepreneurship Theory and Practice, 38(3), pp.671-690.

\bibitem{barab2002evolsocnetw}Barabási, A. L., Jeong, H., Néda, Z., Ravasz, E., Schubert, A., Vicsek, T. (2002) Evolution of the social network of scientific collaborations. Physica A, 311: 590-614.

\bibitem{barabrek1999emergence}Barabási, A.-L., Reka, A. (1999) Emergence of scaling in random networks. Science, 286: 509-12.

\bibitem{wangchen2003complexnetw}Wang, X.F. and Chen, G., 2003. Complex networks: small-world, scale-free and beyond. IEEE circuits and systems magazine, 3(1), pp.6-20.

\bibitem{powkop1996interorg}Powell, W. W., Koput, K. W., Smith-Doerr, L. (1996) Interorganizational collaboration and the locus of innovation: networks of learning in biotechnology. Administrative Science Quarterly, 41: 116-45.

\bibitem{holmeedling2004structime}Holme, P., Edling, C. R., Liljeros, F. (2004) Structure and time evolution of an internet dating community. Social Networks, 26: 155-74.

\bibitem{powwhite2005netwdynam}Powell, W. W., White, D., Koput, K. W., Owen-Smith, J. (2005) Network dynamics and field evolution: The growth of interorganizational collaboration in the life sciences. American Journal of Sociology, 110: 1132-205.

\bibitem{walkkog1997soccapital}Walker, G., Kogut, B., Shan, W. (1997) Social capital, structural holes and the formation of an industry network. Organization Science, 8: 109-25.

\bibitem{hite2001evolfirm}Hite, J.M. and Hesterly, W.S., 2001. The evolution of firm networks: From emergence to early growth of the firm. Strategic management journal, 22(3), pp.275-286.
\bibitem{cookbosch2011handbook}Cooke, P., Asheim, B., Boschma, R., Martin, R., Schwartz, D. and Todtling, F. eds., 2011. Handbook of regional innovation and growth. Edward Elgar Publishing.

\bibitem{gluck2007econgeogr}Glückler, J., 2007. Economic geography and the evolution of networks. Journal of Economic Geography, 7(5), pp.619-634.

\bibitem{sorestu2001syndicat}Sorenson, O., Stuart, T. E. (2001) Syndication networks and the spatial distribution of venture capital investments. American Journal of Sociology, 106: 1546–88.

\bibitem{powkop2002spatclustering}Powell, W. W., Koput, K. W., Bowie, J. I., Smith-Doerr, L. (2002) The spatial clusering of science and capital: accounting for biotech firm-venture capital relationships. Regional Studies, 36: 291-306.

\bibitem{owpow2004knownetw}Owen-Smith, J., Powell, W. W. (2004) Knowledge networks as channels and conduits: The effects of spillovers in the Boston biotechnology community. Organization Science, 15: 5-21.

\bibitem{schutstam2003evolnatyoungfirm}Schutjens, V., Stam, E. (2003) The evolution and nature of young firm networks: a longitudinal perspective. Small Business Economics, 21: 114-34.

\bibitem{has2005unlockregecon}Hassink, R. (2005) How to unlock regional economies from path dependency? From learning region to learning cluster. European Planning Studies, 13: 521-35.
\bibitem{burt2004structholes}Burt, R. S. (2004) Structural holes and good ideas. American Journal of Sociology, 110: 349-99.
\bibitem{mcphers1992socnetworks}McPherson, J. M., Popielarz, P. A., Drobnic, S. (1992) Social networks and organizational dynamics. American Sociological Review, 57: 153-70.

\bibitem{baum2003wheresmworlds}Baum, J. A., Shipilov, A. V., Rowley, T. J. (2003) Where do small worlds come from? Industrial and Corporate Change, 12: 697-725.

\bibitem{row2005timebreakup}Rowley, T. J., Greve, H. R., Rao, H., Baum, J. A. C., Shipilov, A. V. (2005) Time to break up: Social and instrumental antecedents of firm exits from exchange cliques. Academy of Management Journal, 48: 499-520.

\bibitem{dietz1990evolutionary}Dietz, T., Burns, T.R. and Buttel, F.H., 1990, June. Evolutionary theory in sociology: An examination of current thinking. In Sociological Forum (Vol. 5, No. 2, pp. 155-171). Kluwer Academic Publishers-Plenum Publishers

\bibitem{boschmart2007constructing}Boschma, R. and Martin, R., 2007. Constructing an evolutionary economic geography.

\end{thebibliography}
\end{document}